# EFFICIENCY OF PLASMOCHEMICAL PRODUCTION OF HYDROGEN FROM PROPANE UNDER THE INFLUENCE OF LASER RADIATION

Yu.S. Tveryanovich*, A.V. Povolotskiy, S.S. Lunkov

St. Petersburg State University, Institute of Chemistry

198504, St. Petersburg, Universitetsky pr., 26

*e-mail: tys@bk.ru

The plasma-chemical process of producing hydrogen from propane under the excitation by laser radiation has been studied. The study was carried out using femtosecond (35 fs) and nanosecond (7 ns) pulsed laser radiation sources. Experimental dependences of the volumetric hydrogen content in the gas mixture at the outlet of the reactor were measured depending on the propane supply rate. Equations have been proposed that describe the amount of hydrogen produced, taking into account the propane supply rate and the effect of its mixing with the resulting hydrogen. The resulting equations can be applied to the plasma-chemical decomposition of other hydrocarbons. Their application to experimentally studied processes made it possible to calculate the maximum possible efficiency of hydrogen production for the given characteristics of laser radiation. The results obtained made it possible to give recommendations on changing the parameters of laser radiation in order to increase the efficiency of the plasma-chemical process.

**Introduction**

The task of producing hydrogen fuel from natural gas is becoming increasingly urgent. Its solution will make it possible to continue exploitation of the gas fields in the face of stricter environmental requirements. One of the ways to produce hydrogen free of greenhouse gas emissions is the plasma-chemical decomposition of natural gas. Its advantage is the ability to produce hydrogen as needed. Switching the process on and off is easy to implement and does not involve additional energy costs. This means that plants can be located directly next to the consumer, thereby avoiding the problems of transporting and storing hydrogen. Such installations do not require high performance. One way to create plasma is to use laser radiation. Laser-initiated plasma-chemical processes of hydrocarbon dissociation are the subject of close attention of researchers [1-6]. Processes initiated by both nanosecond pulses [4,6] and femtosecond pulses [1–3,5] are considered.

The decomposition of hydrocarbons in laser plasma has a number of features. One of them is a significant number of independent process parameters. On the one hand, this is an advantage, since it allows it to be optimized in many respects. On the other hand, multiparameter optimization is complex and requires the development of approaches that simplify it. The purpose of the work is a comparative study of the production of hydrogen by plasma-chemical decomposition of propane using two types of laser radiation: femtosecond and nanosecond pulse durations with comparable light flux power densities. The research objective also includes the development of model concepts and equations describing the dependence of the amount of hydrogen produced on the rate of propane pumping through the reactor, their application to the obtained experimental data to calculate the maximum possible amount of hydrogen produced at fixed parameters of laser radiation.

**Materials and methods.**

*Propane.* HP Propane (99.9%) supplied by BK Group LLC was used in the work.

*Supplying propane to the reactor and measuring the hydrogen content at the outlet.* Gaseous hydrocarbons were passed through an optical flow cell at a set rate in the range of 1-16 l/h. The propane entering the cuvette is measured using a Swagelok VAF-G2-01m-1-0 rotameter, which allows measuring gas flow with an accuracy of 1 l/h. The volumetric concentration of hydrogen after the flow cell was recorded using an AVP-01G hydrogen analyzer with an accuracy of 0.001%.

Initiation of laser plasma and measurement of its emission spectra

To initiate laser plasma with ultrashort pulses, we used an original setup based on an Astrella femtosecond amplifier (Coherent) with a lasing wavelength of 800 nm, a pulse duration of 35 fs, a pulse energy of 60 µJ, and a pulse repetition rate of 1 kHz. A SpitLight-High-Power 2000-50 laser (Innolas) with a radiation wavelength of 1064 nm, a pulse duration of 7 ns, a pulse energy of 1 J, and a pulse repetition rate of 10 Hz was used as a source of nanosecond laser pulses. Femtosecond pump pulses are focused in the volume of a flow gas cell using a 40x mirror lens, the focal spot diameter is about 4 µm2. Nanosecond laser pulses are focused into the volume of a gas flow cell using a lens with a focal length of 15 cm, which provides the same power density as when focusing femtosecond pulses with a 40x lens. Plasma radiation was directed using a condenser into a single monochromator M-266 (Standa) and recorded by a Hamamatsu H9305-03 photomultiplier tube (PMT). When recording emission spectra, the PMT signal was sent to an analog-to-digital converter (ADC). The parameters of the two types of laser pulses used to initiate laser plasma are given in Table. 1.

Table 1. Parameters of femtosecond and nanosecond laser pulses.

| Pulse type | Wavelength (nm) | Repetition frequency (Hz) | Pulse energy (µJ) | Duration (ps) | Pulse power (GW) | Average radiation power (W) |
|---|---|---|---|---|---|---|
| femto | 800 | 1000 | 60 | 0.035 | 1,7 | 0,060 |
| nano | 1064 | 10 | $10^6$ | 7000 | 0,143 | 10 |

The composition of the mixture leaving the cuvette is determined by the amount of propane entering it per unit time - $x$, and the amount of hydrogen generated per unit time - $h$. Both quantities are measured in $cm^3/s$. The energy of laser pulses is constant, therefore $h$, to a first approximation, is constant. We assume that the components are mixed until completely homogeneous and the system reaches dynamic equilibrium at the moment of registration. After each change in $x$, before measuring $h$, a long time passes, sufficient for the equilibrium concentration to be established. The last condition is the most difficult, since if the generation of hydrogen exceeds the amount of propane supplied, then hydrogen will begin to accumulate, and the system will very slowly come to equilibrium.

The composition of the output mixture, taking into account the decomposition of part of the propane in mole or volume fractions, is equal to $[(x-0{,}25h)_{prop}+ h_{hydr}]$. Average hydrogen fraction by volume of the cuvette:

$$y = \frac{h}{x+0{,}75h} \qquad (1)$$

Let us describe the generation of hydrogen in a propane-hydrogen gas mixture. Obviously, as the hydrogen concentration in the cell increases, the rate of its generation will decrease in proportion to its volume fraction due to a decrease in the propane concentration:

$$h = h_0(1 - y) \qquad (2)$$

Where $h_0$ is the rate of hydrogen generation in pure propane.

From these two equations, neglecting $y^2$ versus $y$, we get:

$$y = \frac{h_0}{x+1{,}75h_0} \qquad (3)$$

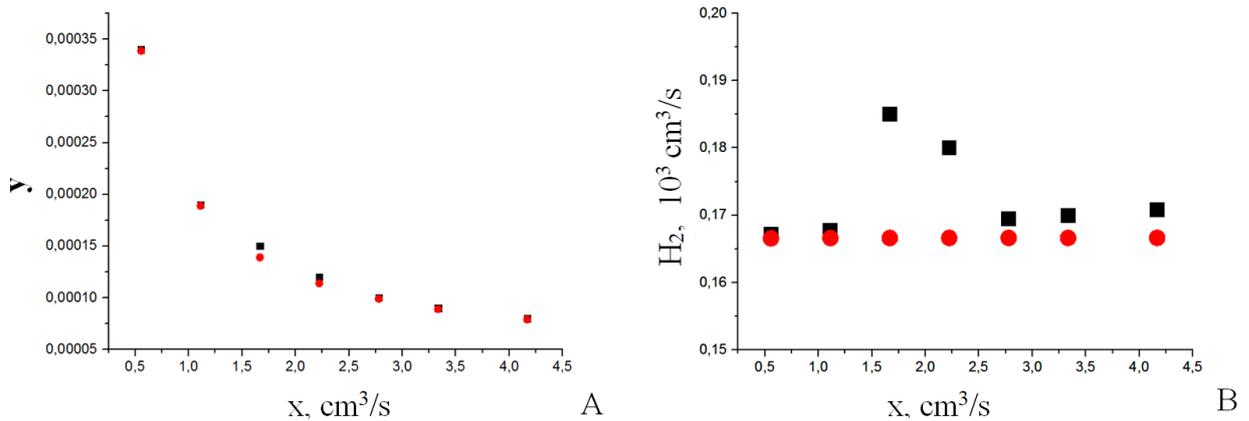

Fig. 1. Generation of hydrogen by femtosecond pulses. A – dependence of the volume fraction of hydrogen in the gas mixture leaving the reactor on the propane pumping rate (squares – experimental values, circles – approximation results). B – dependence of the amount of hydrogen generated per unit time on the rate of propane pumping (squares – experimental values, circles – calculation results).

To take into account possible errors in determining the zero hydrogen concentration when approximating the experimental graph (Fig. 1A), a constant term (A) was added to the right side of equation (3). The equation took the following form:

$$y = A + \frac{h_0}{x + 1{,}75h_0}$$

According to the approximation results, $A = 4.3*10^{-5}$. The $h_0$ parameter calculated from the graph is equal to $0.17*10^{-3}$ cm³/s. By multiplying the experimentally found hydrogen concentration y by the propane pumping speed ($X$), we find the dependence of the hydrogen generation rate ($V$) on the pumping speed (Fig. 1 B). It can be seen that $V$ does not depend on $X$ and is close to $h_0$ (two experimental points that fell out of the general pattern deviate from the value of $h_0$ by no more than 10%). This means that during the short duration of the pulse, the amount of hydrogen produced is small and does not significantly reduce the efficiency of its generation. Therefore, the chosen approximation is correct.

A different situation arises when hydrogen is generated under the influence of pulses with a duration of 7 ns. This duration is $2*10^6$ times longer than the duration of femtosecond pulses. Therefore, the possibility of hydrogen accumulation in the plasma zone is much higher. To confirm this, let us compare the emission spectra of plasma when it is excited by femtosecond and nanosecond pulses (Fig. 2).

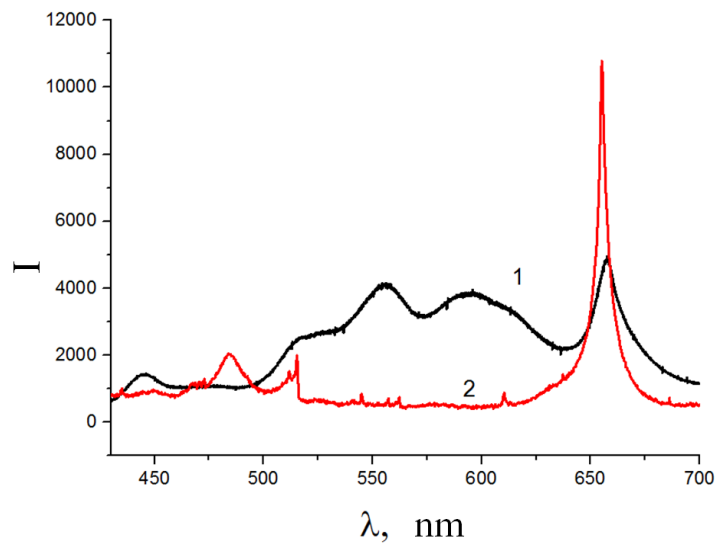

Fig. 2. Spectra of plasma glow under the influence of femtosecond (1) and nanosecond pulses (2).

The assignment of spectral bands was carried out on the basis of literature data [7-10]. The band near 445 nm is attributed to the methylidine radical C-H. The bands at 515, 560, 590 and 615 nm are attributed to diatomic unstable carbon C=C. In the given spectra, the band near 656.3 nm belongs to the α-transition in the Balmer series (see, for example, [11]) for single hydrogen atoms – the transition between the second and first excited states. The β transition in the Balmer series is located at 486.1 nm. When going from the spectrum of a plasma excited by a femtosecond pulse to the spectrum of a plasma excited by a nanosecond pulse, a significant (by an order of magnitude) increase in the contribution of spectral lines associated with hydrogen to the total intensity of the plasma glow is observed. This confirms the assumption that increasing the pulse duration leads to the accumulation of hydrogen in the region of plasma existence. Let us take this circumstance into account in further calculations, especially since equation (3) does not allow us to satisfactorily approximate the experimental data for the dependence of the hydrogen concentration at the outlet of the reactor on the propane supply rate in the case of nanosecond excitation of the plasma.

If a uniform distribution of hydrogen over the volume does not have time to establish itself, then we will assume that the effective concentration of hydrogen displacing propane in the plasma-chemical reaction zone is greater than the average concentration over the volume by factor $C$. Then in equation (2) the factor $C$ will appear in front of $y$. Having solved the new system equations we get

$$y = \frac{h_0}{x + h_0(0{,}75 + C)}. \quad (4)$$

In this case $(C^*y) \leq 1$. Similar to the case of femtosecond pulses, the constant $A$ was also introduced into the right side of equation (4). We use this equation to describe the dependence of the fraction of hydrogen in the gas mixture leaving the reactor on the propane pumping rate for the case of laser pulses with duration of 7 ns.

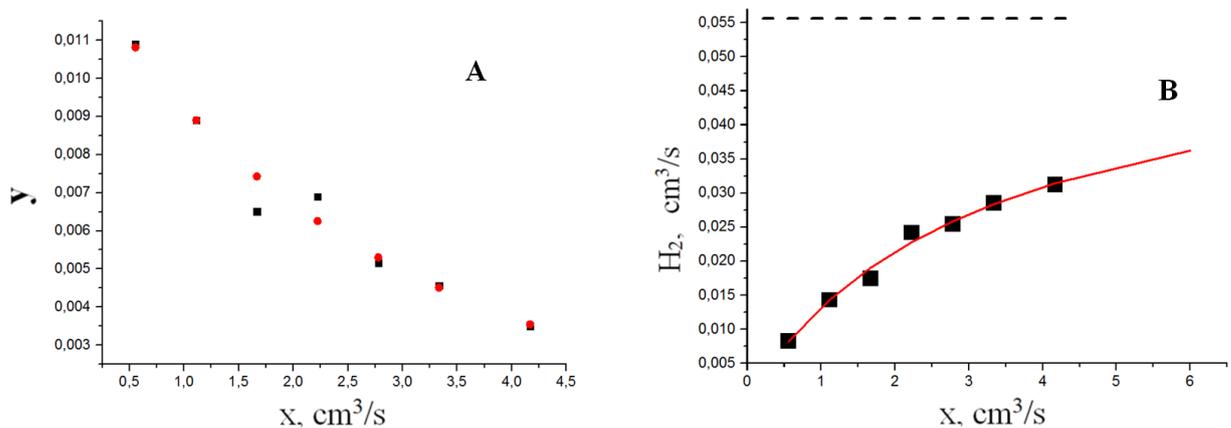

Fig. 3. Generation of hydrogen by nanosecond pulses. A – dependence of the volume fraction of hydrogen in the gas mixture leaving the reactor on the propane pumping rate (squares – experimental values, circles – approximation results). B – dependence of the amount of hydrogen generated per unit time on the rate of propane pumping (squares – experimental values, line – calculation result, dotted line – $h_0$ value).

According to the results of approximation of experimental data using equation (4) (see Fig. 3A), the following results were obtained: $A = 0$, $h_0 = 0.022$ cm³/s, which is 13 times higher than in the case of femtosecond excitation. Parameter C is equal to 66. Parameter C is of great importance, which indicates a significant accumulation of hydrogen in the plasma region compared to its average concentration throughout the reactor volume. However, the condition $(C^*y) \leq 1$ is satisfied.

Using the found parameters, we calculate the dependence of the amount of hydrogen leaving the reactor per unit time on the propane supply rate. It, as noted above, is equal to the product $(y\,x)$. In this case, experimental values of the volume fraction of hydrogen in the gas mixture at the exit from the reactor are used, considering correction $A$ (points in Fig. 3 B). The line represents the calculated values of the amount of hydrogen generated per second. Description of this dependence using equation (4) and the found parameters allows, by extrapolation, to find the rate of propane pumping at which the amount of hydrogen produced will approach the limit value ($h_0$ = 0.022 cm$^3$/s) and will differ from it by 5%. This speed is equal to $V_c$ = 60 cm$^3$/s. Such a large value of $V_c$ is the result of the accumulation of hydrogen in the plasma region during a nanosecond pulse and the need to remove it from this region during the time between pulses. In the case of femtosecond pulses, as can be seen from Fig. 1, $V_c$ < 0.5 cm$^3$/s, since in the entire studied range of propane supply rates ($V_c$) $h_0$=0.17*10$^{-3}$ cm$^3$/s.

In a number of works [12-15], to analyze the energy efficiency of hydrogen production, storage and transportation, the energy profitability criterion EROEI (Energy Returned on Energy Invested) is used, determined by the formula: EROEI = $E_o/E_s$, where $E_o$ is the amount of energy obtained when using energy resource; $E_s$ is the amount of energy expended to obtain the energy resource used. In our case, the interest is in the maximum possible energy efficiency of the process of producing hydrogen as a result of the decomposition of hydrocarbons in laser plasma under a given laser operating mode. Therefore, by analogy with EROEI, we will use the energy efficiency of this process in the form:

$$f = \frac{h_0 \cdot T}{W} \qquad (5)$$

$W$ - is the power of the light flux (W), $T$ - is the calorific value of hydrogen (J/cm3) [16]. Substituting into equation (5) the found values of $h_0$ and the values of $W$ from the table. 1, we find that for femtosecond radiation $f$ = 3.3%, and for nanosecond radiation – 2.6%.

Thus, an increase in the duration of the laser pulse at a sufficiently high power can lead to an increase in the hydrogen concentration in the laser plasma region in comparison with the average over the reactor volume. The result of this is a decrease in the efficiency of hydrogen production. The proposed method allows us to take this effect into account. The use of this method is useful in searching for an effective laser radiation mode for producing hydrogen by decomposition of low molecular weight hydrocarbons.

*Research funding.* The research was supported by a joint grant from the Russian Science Foundation No. 22-23-20038. https://rscf.ru/project/22-23-20038/ and the St. Petersburg Science Foundation in accordance with agreement dated April 14, 2022 No. 34/2022. The measurements were carried out in the Science Park of St. Petersburg State University (resource center Optical and Laser Methods for Matter Studying).

"The authors declare that there are no conflicts of interest to disclose in this article."

Information about authors.

Tveryanovich Yuri Stanislavovich, ORCID: http://orcid.org/0000-0003-4343-9817

Povolotskiy Alexey Valerievich, ORCID: http://orcid.org/0000-0001-7931-9572

Lunkov Svyatoslav Sergeevich, ORCID: http://orcid.org/0009-0000-5897-4455